\documentclass[traditabstract]{aachanged}
\usepackage[english]{babel}
\usepackage{float}
\usepackage{longtable}
\usepackage{graphicx}	
\usepackage{epsfig}
\usepackage{epsf}
\usepackage{amsmath}	
\usepackage{amssymb}	
\usepackage{xcolor,soul}
\usepackage{subcaption}
\usepackage{hyperref}
\definecolor{darkblue}{RGB}{0, 0, 200}
\hypersetup{
    colorlinks=true,
    linkcolor=blue,
    filecolor=magenta,      
    urlcolor=darkblue,
    }
\usepackage{tabu}
\usepackage{soul}
\usepackage[normalem]{ulem}

\begin{document}

\title{\bf 
Low-Energy Neutrinos from Primordial Black Holes: \\
A New Possibility for Observing Hawking Radiation}
\author{Yu.A. Lysyy$^{1}$, P.A. Kislitsyn$^1$\thanks{E-mail: pavel.kislitsyn@gmail.com}, A.V. Ivanchik$^1$}
\authorrunning{Lysyy et al.}
\titlerunning{Low-Energy Neutrinos From PBH}
\date{$\qquad\qquad\qquad\qquad\qquad\qquad\qquad\qquad\qquad$Accepted December 12, 2024\\ DOI: 10.1134/S1063773725700021}

\institute{\centering \it{$^{1}$Ioffe Institute, Russian Academy of Sciences,
Saint Petersburg, 194021 Russia}}

\abstract{The study of primordial black holes and the Hawking radiation they can produce represents an important step in understanding the role of these phenomena in the cosmological evolution of the Universe. Primordial black holes can be part of dark matter, embryos of supermassive black holes, and sources of Hawking radiation, which, unlike the radiation of other black holes, can be observable. At the same time, in the evolution of the Universe from the Big Bang to the present day, primordial black holes lose most of their mass in the form of neutrino radiation. This happens because for black holes with $M<10^{23}$\,g, along with the radiation of massless particles, the radiation of the lightest massive particle -- the neutrino -- is added. This radiation turns out to be dominant, and since by the present moment ($t_0=13.8$ billion years) only black holes with masses $\lesssim 10^{15}\,$g have largely evaporated, the total emission spectrum of primordial black holes is dominated by the neutrino component. In this paper, we present new estimates of the neutrino spectra emitted by primordial black holes of different masses, focusing for the first time on the low-energy ($E_{kin} \in [0.01 \div 1]\, $eV) radiation. The calculations show that black holes in the mass range $[10^{9}\div10^{11}]\,$g emit neutrinos with intensities exceeding the background fluxes from known astrophysical sources in the low-energy range, while in the high-energy range the emission will be under the background without contradicting observational constraints. These results open new possibilities for the potential observation of primordial black hole radiation and may stimulate the development of neutrino detection technologies in the low-energy range. The observation of neutrinos in this range is one of the few opportunities to confirm the existence of Hawking radiation.}


\maketitle

\section{Introduction}

The standard cosmological model currently describes most of the observed phenomena occurring on the scale of the Universe. However, some questions remain relevant, and one of them is related to the possibility of the existence of primordial black holes, since it is for them that the radiation predicted by Hawking (\hyperlink{refs:27}{Hawking, 1974}; \hyperlink{refs:28}{Hawking, 1975}) can be detected, whereas for stellar-mass black holes and supermassive black holes this radiation is unobservable with currently available technology.
In addition, primordial black holes may be a constituent part of dark matter, as well as being the embryos of supermassive black holes at the centers of galaxies (for details, see, e.g., the review by \hyperlink{refs:12}{Carr et al., 2021}). The question of the generation of baryon asymmetry by primordial black holes is also of interest -- another problem of modern cosmology (\hyperlink{refs:8}{Dolgov \& Pozdnyakov, 2021}). 

The existence of black holes of solar ($[1\div 100]M_{\odot}$) and intermediate ($[10^2\div 10^5]M_{\odot}$) masses, as well as supermassive ($[10^5\div 10^{11}]M_{\odot}$) black holes is almost beyond doubt today. The collapse of massive stars leads to the formation of black holes of solar masses, the merger of black holes in double systems provides the formation of black holes of intermediate masses, which has recently been detected in the registration of gravitational waves (as a review of recent data, see, for example, the work of the \hyperlink{refs:19}{LIGO Scientific Collaboration, 2024}). Only the question of mass accretion by supermassive black holes remains not fully understood. According to modern ideas, the existence of supermassive black holes at large redshifts, when the age of the Universe was $\lesssim 1$ billion years, cannot be explained by the formation of solar-mass black holes and their subsequent growth, however, observations show the presence of supermassive black holes in the Universe at redshifts $z\sim 6-7$. Thus, in \hyperlink{refs:1}{Eilers et al., 2023} , an estimate of the mass of a supermassive black hole is presented, which is 
$M\approx 10^{10} M_{\odot}$, toward quasar J0100+2802 at redshift $z = 6.327$, corresponding to an age of the Universe of only 800 Myr. The paper by \hyperlink{refs:31}{Yang et al., 2021} presents a set of 37 quasars at comparable redshifts where black holes have masses in the range $[0.3\div3.6]\times 10^{10} M_{\odot}$.
One possible explanation for the existence of supermassive black holes in such an early Universe would be primordial black holes as embryos (see, e.g., \hyperlink{refs:1}{Eilers et al., 2023}).

Primordial black holes, among all others, are singled out as a separate problem because observational evidence for their existence has not yet been found, and the mechanisms of their formation contain significant uncertainties and depend on the physical conditions in the environment where they were formed. It is generally believed that a primordial black hole of mass $M$ may have formed in the early Universe when the mass enclosed within the cosmological horizon was comparable to $M$. \hyperlink{refs:12}{Carr et al., 2021}, describe several possible ways in which primordial black holes could have formed from matter within the cosmological horizon at a particular point in time. For example, black holes with mass $\sim 10^{10}\,$g could have formed at times $\sim 10^{-28}$ sec when the plasma temperature was $T \sim 10^{11} $ GeV/$k$. Thus, the mass spectrum of primordial black holes can be quite broad, and the Hawking radiation of black holes with masses from $M_{\rm Pl} \approx 10^{-5}$ g to $10^{15}$ g can appear in observations. At the same time, primordial black holes of higher masses lose almost no mass over the lifetime of the Universe via Hawking radiation (see Tab.~1).The possibilities of detecting Hawking radiation are discussed in detail, e.g. in \hyperlink{refs:12}{Carr et al., 2021}, with these studies focusing on observations in the high-energy range ($E_{kin} \gtrsim 1 \,$MeV). In our work, we focus on the low-energy range of $E_{kin} \in [0.01 \div 1]\, $eV, where Hawking radiation in neutrinos can exceed the background fluxes without conflicting with other observations. 

The neutrino is another phenomenon attracting the interest of physicists today. This is mainly due to the amazing properties of neutrinos, which may require the construction of a theory beyond the Standard Model. Many ambitious experiments to detect neutrinos in different energy ranges are in preparation. The \hyperlink{refs:13}{JUNO Collaboration, 2016}; the \hyperlink{refs:20}{Hyper-Kamiokande Proto-Collaboration, 2018}; the \hyperlink{refs:14}{DUNE Collaboration, 2020}, aim to study neutrinos with energies of $\gtrsim 1 $MeV, and the \hyperlink{refs:17}{PTOLEMY Collaboration, 2019}, aims to detect the signal of cosmological neutrinos with energies of $\sim 10^{-3}$ eV. Of particular interest in recent years has been the study of so-called sterile neutrinos. They do not interact with ordinary matter, but can oscillate into Standard Model neutrinos and thus have a significant impact on estimates of cosmological parameters (see, e.g., \hyperlink{refs:30}{Chernikov \& Ivnachik, 2022}; \hyperlink{refs:9}{Ivanchik et al., 2024a}). The question of the existence of sterile neutrinos, however, cannot be completely closed today, since experimental results are in poor agreement with each other and sometimes completely contradict each other. For example, \hyperlink{refs:23}{Serebrov et al., 2021} indicate a reliable confirmation of the existence of a fourth type of neutrino, with which the results of \hyperlink{refs:4}{Barinov et al., 2021}, are in agreement, but the \hyperlink{refs:18}{STEREO Collaboration, 2023}, and the \hyperlink{refs:16}{PROSPECT Collaboration, 2024}, report that the hypothesis of the existence of light sterile neutrinos can be rejected. Future independent experiments will help to clarify this controversial situation.

\section{Mass spectrum of primordial black holes}

\noindent
At formation of primordial black holes in the first moments in the lifetime of the Universe they could have different masses. The mass distribution of black holes is determined by the model of their formation and affects the shape of the total spectrum of their radiation. The lack of observational evidence of radiation allows one to limit the set of acceptable mass spectra, as well as the contribution of primordial black holes to dark matter. This was first done in the works by \hyperlink{refs:26}{Hawking, 1971}, \hyperlink{refs:29}{Chapline, 1975}, and was later done by \hyperlink{refs:6}{Dolgov \& Silk, 1993}, who, among others, showed that the mass spectrum of primordial black holes in their model is log-normal. Since then, the abundance constraints of primordial black holes and their mass spectrum have been actively investigated (see, e.g., \hyperlink{refs:21a}{Novikov et al., 1979}, \hyperlink{refs:11}{Carr et al., 2010}; \hyperlink{refs:12}{Carr et al., 2021}). Distributions such as lognormal, power-law, and monochromatic are commonly used to describe the mass spectrum of primordial black holes. To simplify the calculations, in our work we studied the monochromatic distribution, which is expected to bias the result only quantitatively, but correctly reproduces all features at the qualitative level.

The most common mass spectra of primordial black holes used in the literature (monochromatic, power-low and log-normal):  
\begin{align*}
    \phantom{}\dfrac{dn}{dM} &= n_{\mathrm{PBH}} \delta(M - M_0)\phantom{\dfrac{1_1}{2_2}} &\\
    \dfrac{dn}{dM} &= n_{\mathrm{PBH}} \dfrac{\gamma - 1}{M_{\mathrm{max}}^{\gamma - 1}-M_{\mathrm{min}}^{\gamma - 1}}  M^{\gamma-2},  \ \ (M_{\mathrm{min}} \leq M \leq M_{\mathrm{max}}) & \\
    \dfrac{dn}{dM} &= n_{\mathrm{PBH}}\dfrac{1}{\sqrt{2\pi\sigma^2}M} \exp\left(-\dfrac{\ln^2(M/M_{BH})}{2\sigma^2}\right),  \ \  & \\
\end{align*}
where $n_{\mathrm{PBH}}$ -- number density of primordial black holes and $M_{0}, M_{\mathrm{min}}, M_{\mathrm{max}}, \gamma, 
M_{BH}, \sigma$ -- parameters of distributions.

Where $n_{\mathrm{PBH}}$ is the number density of primordial black holes, and $M_{0}, M_{\mathrm{min}}, M_{\mathrm{max}}, \gamma, 
M_{BH}, \sigma$ -- parameters of the distributions.




\definecolor{c1}{RGB}{0,0,128}
\definecolor{c2}{RGB}{21,21,144}
\definecolor{c3}{RGB}{42,42,162}
\definecolor{c4}{RGB}{64,64,178}
\definecolor{c5}{RGB}{85,85,195}
\definecolor{c6}{RGB}{106,106,212}
\definecolor{c7}{RGB}{128,128,230}
\definecolor{c8}{RGB}{50,205,50}
\definecolor{c9}{RGB}{255,128,128}
\definecolor{c10}{RGB}{249,96,96}
\definecolor{c11}{RGB}{242,64,64}
\definecolor{c12}{RGB}{236,32,32}
\definecolor{c13}{RGB}{230,0,0}


\begin{table*}[t]
\caption{Characteristics of different black holes and their radiation\vspace{1mm}}
\label{tab:BH_examples}
\centering 
\vspace{-4mm}
\begin{tabu}{ | c | c | c | c |}
\hline
\rule{0pt}{2.4ex}
 Mass & Schwarzschild radius & Temperature & Time of evaporation$^1$ \\ 
\hline \hline
\rowfont{\color{c1}}
\rule{0pt}{2.4ex}
$10^9M_{\odot}$ & $20 \ $a.u. & $6 \times 10^{-17} $\ K & $\sim 10^{82} \, t_0$ \\
\rowfont{\color{c2}}
 $4\times 10^6M_{\odot}$ & $0.08 \ $a.u. & $1 \times 10^{-14} $ \ K & $\sim 10^{75}\, t_0$ \\
\rowfont{\color{c3}} 
$\mathbf{M_{\odot}}$  & $\textbf{3 km} $  & $\mathbf{6 \times 10^{-8} \ K}$  & $\mathbf{\sim 10^{55}\, t_0}$ \\  
\rowfont{\color{c4}}
$M_{\oplus}\approx 6\times 10^{27}  $g & $0.9 \, \text{cm}$ & $0.02  $\ K & $\sim 10^{38} \,t_0$ \\  
\rowfont{\color{c5}}
$ 4.5\times 10^{25}  $ g $ ^{\dagger}$ & $67 \, \mu m$ & $~~~~~~2.7255  $\ K $\left(T^0_{\rm CMB}\right)$& $\sim 10^{32} \,t_0$ \\  
\rowfont{\color{c6}}
$10^{23} $\ g & $0.1 \, $ $\mu m$ & $1\times10^3 $\ K & $\sim 10^{24} \, t_0$ \\ 
\rowfont{\color{c7}}
$2\times10^{16} $\ g $ ^{\dagger\dagger}$ & $20 \, $ fm & $5\times10^9 $\ k & $\sim 10^{4} \, t_0$ \\ 
\rowfont{\color{c8}}
$\phantom{\dfrac{1}{2}}10^{15} $ g$\phantom{\dfrac{1}{2}}$ & $1 \,  $ fm & $1 \times 10^{11} \, $K & $\sim t_0$ \\  
\rowfont{\color{c9}}
$10^{13} $\ g & $1\times10^{-15} $\ cm & $1 $ GeV/k & $\sim 10^{4} \, $years  \\ 
\rowfont{\color{c10}}
$10^{11} $\ g & $1\times10^{-17} $\ cm & $100 $ GeV/k  & $\sim 1 \, $week \\  
\rowfont{\color{c11}}
$10^9 $\ g & $1\times10^{-19} $\ cm & $10^4 $ GeV/k  & $\sim 1 \, $s \\  
\rowfont{\color{c12}}
$10^7 $\ g & $1\times10^{-21} $\ cm & $10^6 $ GeV/k  & $\sim 10^{-6} \, $s \\  
\rowfont{\color{c13}}
\rule[-1.2ex]{0pt}{0pt}
$^*$$M_{\mathrm{Pl}}\sim 10^{-5}$\ g & $^*$$2l_{\mathrm{Pl}}\sim 10^{-33}$\ cm & $^*$$T_{\mathrm{Pl}}/8\pi\sim 10^{19}$ \!GeV/k & $^{**}$$[t_{\mathrm{Pl}}\div \infty]$ \\ 
\hline
\end{tabu}
\flushleft $^{1}$ - order of magnitude estimation of evaporation time according to $\tau \sim t_0 \left(\dfrac{M}{10^{15} {\rm g}}\right)^3$ 
(see, e.g., the work of \hyperlink{refs:5}{Bernal et al., 2022}), where $t_0 \approx 13.8$ Gyr -- the age of the Universe. \\
$^*$ - results of extrapolation of the Hawking radiation theory to the Planck scale are presented. More reasonable values require the use of quantum gravity. \\
$^\dagger$  - In the modern era, for black holes with higher masses ($M > 4.5\times 10^{25} $ g) 
Hawking radiation will be completely compensated by absorption of relic radiation, and they will not only not evaporate, but will gain mass. \\
$^{\dagger\dagger}$ - for black holes with lower masses ($M < 2\times 10^{16} $ g) to the emission of photons and neutrinos radiation of $e^{\pm}$-pairs is added . \\  
$^{**}$ - The Hawking radiation can stop when the minimum mass is reached, forming a remnant -- the so-called Planckion. In case of further evaporation, the black hole will completely evaporate in the Planck time.  \\
$M_{\odot}\approx 2\times10^{33}$ g - Sun mass. \\
$M_{\oplus}$ - Earth mass. \\
$M_{\mathrm{Pl}}, l_{\mathrm{Pl}}, T_{\mathrm{Pl}}, t_{\mathrm{Pl}}$ - Planck mass, length, temperature and time.\\
\end{table*}

\section{Composition and properties of Hawking radiation}

The quantum structure of matter and interactions near the event horizon of a black hole leads to the emergence of the Hawking radiation (\hyperlink{refs:27}{Hawking, 1974}; \hyperlink{refs:28}{Hawking, 1975}) - the quasi-black body radiation of all particles of the Standard Model. The temperature of this radiation is inversely proportional to the black hole mass: 
\begin{equation}
    T = \dfrac{\hbar c^3}{8\pi k G M_{\rm BH}} \approx 6.2 \times 10^{-8}\frac{M_{\odot}}{M_{\rm BH}}, \quad{\rm ~K}
\end{equation} 

Where $M_{\odot}$ is the mass of the Sun, $M_{\rm BH}$ is the mass of a black hole, $\hbar$ is Planck constant, $k$ is Boltzmann's constant, $G$ is the gravitational constant, $c$ is the speed of light. 

The evaporation time of a black hole can be estimated using a formula (see, e.g., \hyperlink{refs:5}{Bernal et al., 2022}):
\begin{equation}
    \tau \sim t_0 \left(\dfrac{M_{\rm BH}}{10^{15} {\rm g}}\right)^3
\end{equation}

Where $t_0\approx 13.8$ Gyr is the modern age of the Universe. It should be noted, however, that this formula can be used only for estimation of the evaporation time by order of magnitude, since it does not take into account the change of the Hawking radiation composition as the black hole evaporates, which is described in detail below.

\begin{figure}[H]
\begin{center}
        \includegraphics[width = 0.5\textwidth]{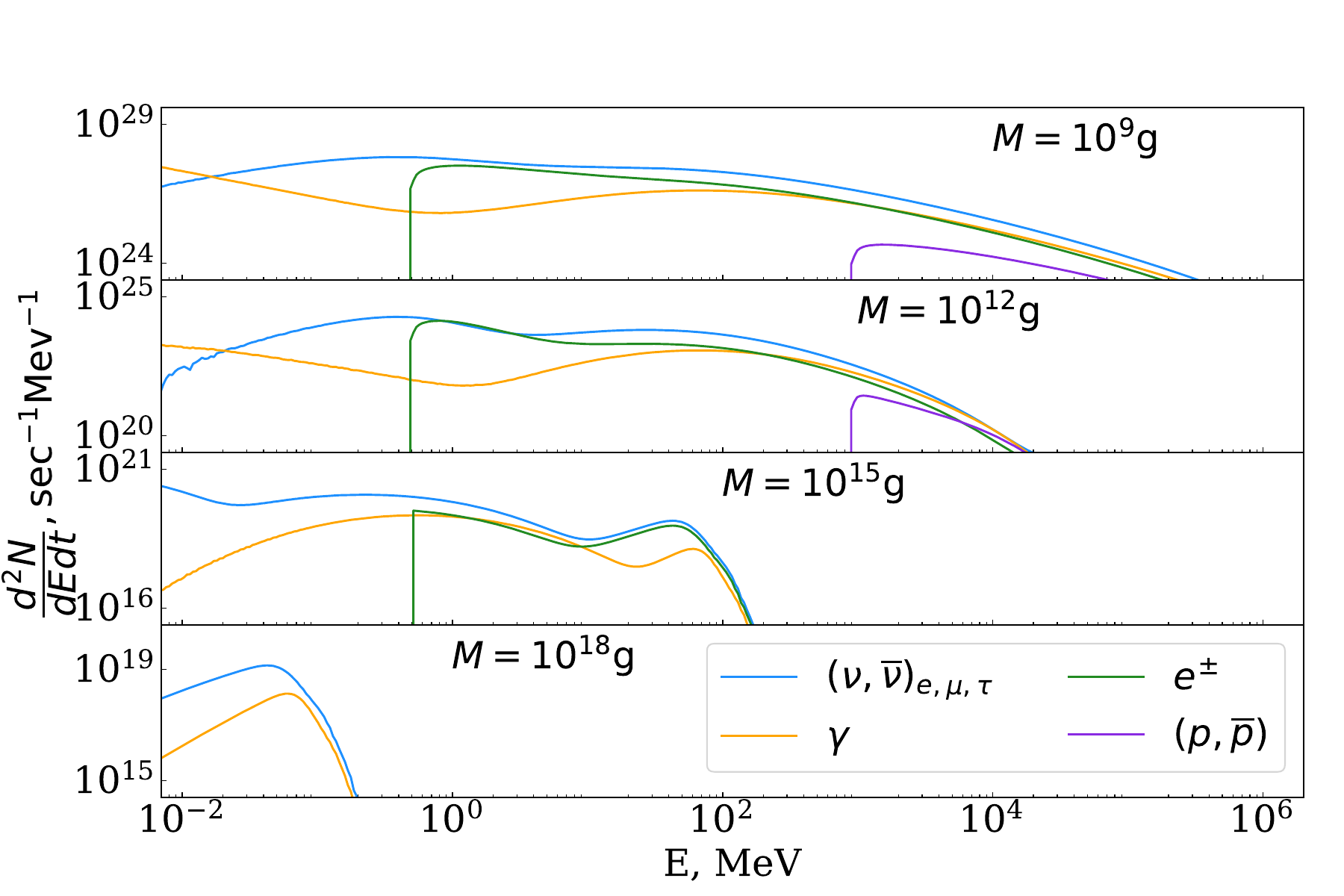}
        \caption{Instantaneous spectra of stable particles from black holes with masses $\left\{10^{9}, 10^{12}, 10^{15}, 10^{18}\right\}$\,g. The blue curve shows the total emission of neutrinos and antineutrinos of all sorts, the orange curve shows the emission of photons, the green curve shows the emission of electrons and positrons, and the violet curve shows the emission of protons and antiprotons.}
        \label{inst}
\end{center}
\end{figure}

According to modern ideas, the evaporation of black holes occurs as follows. Massless particles are emitted by all black holes, with particles with the lowest spin being born most easily. (See, for example, \hyperlink{refs:24}{Harlow, 2016}). In the Standard Model, such particles are photons, but it is worth mentioning the hypothetical gravitational interaction transfer particles -- gravitons. \hyperlink{refs:21}{Page, 1976a}, concluded that for non-rotating stellar-mass black holes, $\approx 2\%$ of the total radiation is emitted via gravitons, with Page taking into account only the electron and muon neutrinos known at that time, which were assumed to be massless. Also \hyperlink{refs:22}{Page, 1976b}, showed that for rotating black holes the composition of the radiation can differ significantly in the direction of increasing the contribution of gravitons, but such black holes lose their momentum much faster than their mass. This allows one to use the non-rotating black hole approximation for calculations. For massive particles, their radiation process has a threshold character, exponentially suppressed at low temperatures. Particles with mass $m_i$ will begin to emit when the size of the black hole decreases to the Compton wavelength of this particle, which is equivalent to the condition $kT \sim m_ic^2$. For neutrinos with mass $0.1$ eV/$c^2$ such a threshold mass of the black hole will be $\sim10^{23}$g, and, for example, for the birth of electrons and positrons the determining mass will be $\sim10^{16}$g (see Tab.~\ref{tab:BH_examples} and Fig.~\ref{inst}). Thus, the next particles after photons, which will be emitted by black holes as their mass decreases, will be neutrinos. At the same time, as it will be shown further, the instantaneous spectra of neutrinos exceed the spectra of photons, a key reason for this is the remarkable superiority of neutrinos, which possess three times more degrees of freedom than photons. 

\noindent
\newpage
\section{Method of calculation of particle emission spectra}

\noindent

The total emission of particles of the variety $i$ has two components:

\begin{equation}
    \frac{d^2N_i}{dEdt}(E_i, T) = \frac{d^2N_i^{pri}}{dEdt}(E_i, T) + \frac{d^2N_i^{sec}}{dEdt}(E_i, T)    
\end{equation}

Where the first term is responsible for the primary radiation, which is directly Hawking radiation, and the second term is the radiation due to the decay of gauge bosons and heavy leptons of the primary radiation, as well as the birth of hadrons from quarks and gluons, called secondary radiation. 

Black holes emit particles during the whole time of evaporation, so the process of radiation production is stretched in time. Taking into account the expansion of the Universe and the distribution of black holes in space, the particle flux registered on the Earth today will be expressed by the formula (\hyperlink{refs:12}{Carr et al., 2021}, \hyperlink{refs:5}{Bernal et al., 2022}):

\begin{equation}
    \frac{d\Phi}{dE_{\nu}}(E_{\nu}) = \frac{c}{4\pi} n_{\rm PBH}^0 \int\limits_{t_{min}}^{t_{max}}dt (1+z) \frac{d^2N}{dtdE_{\nu}}(t, E_{\nu}(1+z))
    \label{eq:3}
\end{equation}

Where $n_{\rm PBH}^0$ -- the Universe average number density of primordial black holes at the moment of formation, multiplied by the factor $(1+z)^{-3}$, describing the expansion of the Universe from the moment of black hole formation up to the present moment, i.e. this value describes the number density of primordial black holes at the present moment in the case when the black hole mass is $\gtrsim10^{15}\,$g, but it has no relation to the present-day number density of primordial black holes of small mass ($\lesssim 10^{15}\,$g).; $t_{min}$ is the starting time of the spectrum summation, taken in this paper to be 1 sec, i.e., the time of neutrino decoupling, as is commonly done (see, e.g., \hyperlink{refs:5}{Bernal et al., 2022}). At earlier times ($t \lesssim 1$ sec, corresponding to a temperature of $T \gtrsim 2\ $\ MeV) the rate of electroweak reactions is higher than the expansion rate of the Universe, because of which neutrinos emitted by black holes quickly thermalize and become part of the background.
The upper limit of integration $t_{max}$ is determined by the time of evaporation of the black hole: for those black holes which have already evaporated by the present moment, $t_{max}$ is equal to the moment of their evaporation, and for black holes which have not evaporated yet, $t_{max}$ is equal to the age of the Universe at the present moment.

The open source code BlackHawk (\hyperlink{refs:2}{Arbey \& Offinger 2019}; \hyperlink{refs:3}{Arbey \& Offinger 2021}) was used to calculate the instantaneous spectra of neutrinos emitted by black holes. It calculates the emission of all Standard Model particles via Hawking radiation and the secondary emission spectra of the particles. The secondary spectra were calculated using the PYTHIA hadronization tables (\hyperlink{refs:25}{Sj\"{o}strand et al., 2015}) used within the BlackHawk code. The PYTHIA tables are calculated for plasma temperatures from $5$ GeV/k to $10^5$ GeV/k and are just right for analyzing the emission of black holes of the studied masses. It is worth noting that in accordance with the Standard Model, the neutrino mass is assumed to be zero when calculating the emission spectra, but for black holes with masses $M \ll 10^{23}\,$g the presence of mass does not affect the neutrino emission spectra. Nevertheless, the presence of neutrino mass can affect the neutrino propagation through the Universe. For example, neutrinos can be captured by the gravitational potential of galaxies. 

\begin{figure}[H]
\begin{center}
        \label{fig:density_constraints}
        \includegraphics[width =0.5\textwidth]{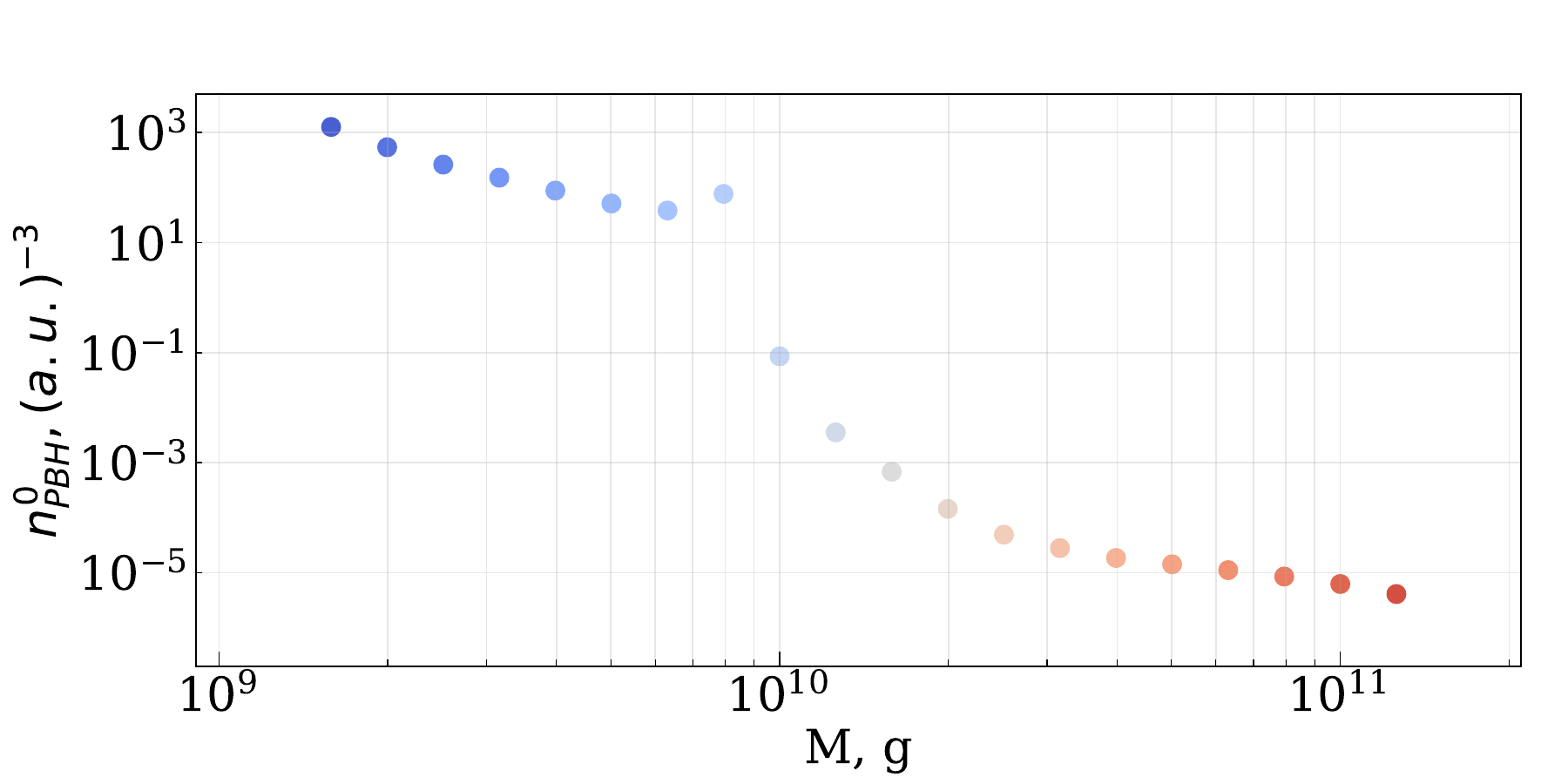}
        \caption{The Universe-averaged number densities of primordial black holes for a set of monochromatic distributions with different masses. The values were calculated using data from \protect\hyperlink{refs:8}{Carr et al., 2021} and are consistent with the existing constraints derived from various astrophysical observations.}
\end{center}
\end{figure}

\begin{figure*}
\begin{center}
        \includegraphics[width = \textwidth]{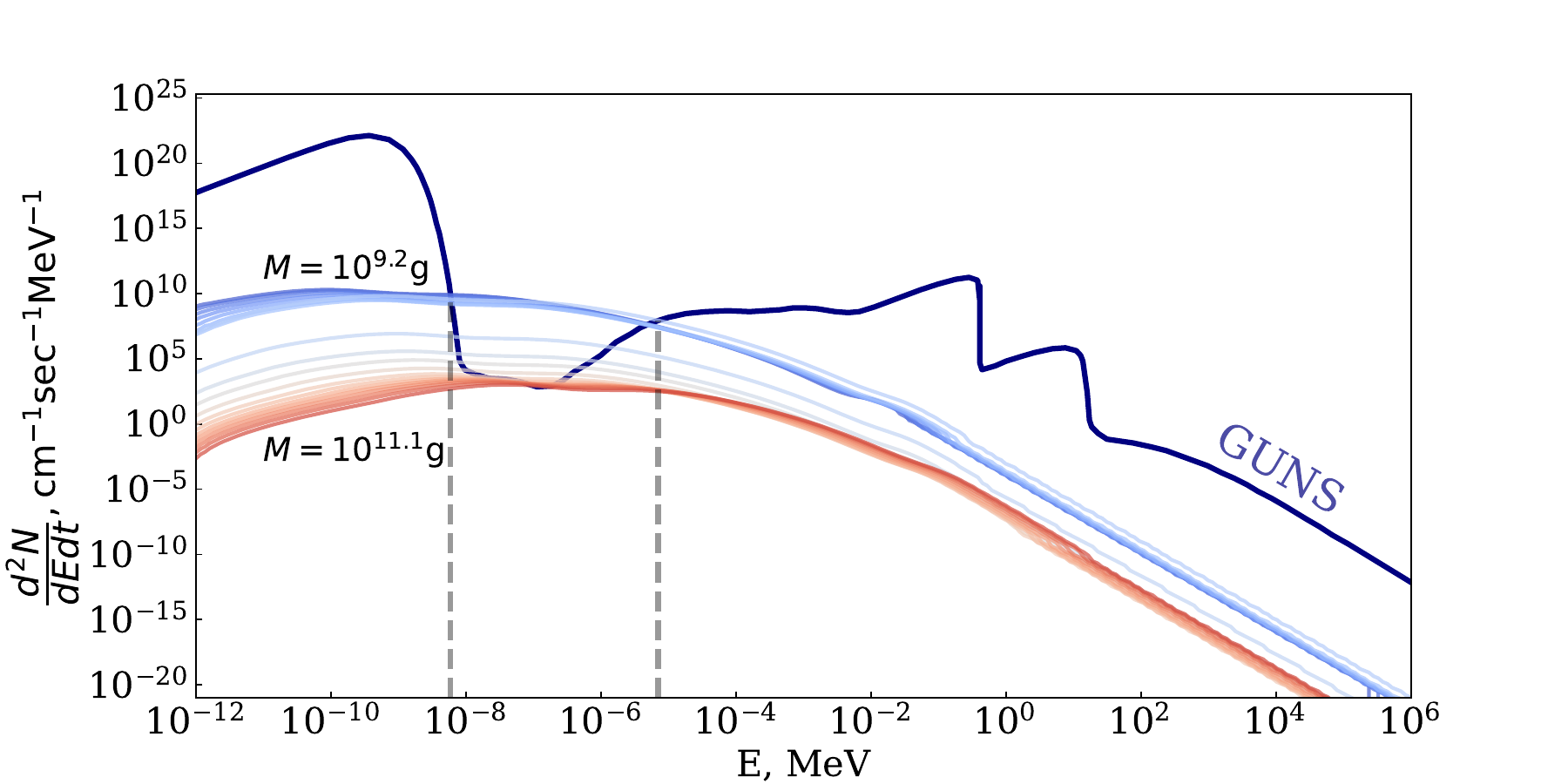}
        \caption{Spectral density of the neutrino flux from primordial black holes reaching an observer on the Earth in the massless neutrino approximation. Thin curves correspond to different monochromatic spectra of black hole masses. Shown are curves for the distributions of black holes with masses in the range $[10^{9.2} \mathrm{\text{\,g}};10^{11.1} \mathrm{\text{\,g}} ]$ with a logarithmic step of $0.1$. The dark blue curve shows the neutrino background (Grand Unified Neutrino Spectrum) due to other sources, taken from \protect\hyperlink{refs:10}{Ivanchik et al., 2024b.}}
        \label{fig:main}
\end{center}
\end{figure*}

The total spectrum, which takes into account the evolution of radiation as the Universe expands, was calculated according to the formula (\ref{eq:3}) using the ``stack.c'' script written in addition to the basic BlackHawk procedures by its authors. The script calculates the radiation assuming no interaction of the emitted particles with other particles as they propagate through the universe. The number density of primordial black holes $n_{\rm PBH}^0$ for each mass was chosen to be as large as possible, while still not contradicting the existing constraints from \hyperlink{refs:12}{Carr et al., 2021}. Fig. \ref{fig:density_constraints} presents the values of the number densities of the primordial black holes studied in this work, determined from the constraints. In the most important range of primordial black hole masses ($[10^9\div 10^{11}]$g, see Fig. \ref{fig:main}), the tightest, and hence determinant constraint is provided by observations of primordial elemental abundances.

\section{Results}

The first important thing to note is that for black holes of the mass range under study, the neutrino emission always exceeds the photon emission. In Fig. \ref{inst} we present the total instantaneous spectra of neutrinos and photons for black holes of different masses. The condition $kT \sim m_{\nu}c^2$ defines the black hole mass boundary, starting from which the effective neutrino emission occurs. The expression of the black hole mass from this condition is: $M_{\rm BH} = \dfrac{M_{\rm Pl}}{8\pi} \left(\dfrac{m_\nu}{M_{\rm Pl}}\right)^{-1}$, where $M_{\rm Pl}$ is the Planck mass. Taking $m_\nu = 0.1$ eV (the order of the upper limit on the sum of neutrino masses from the \hyperlink{refs:15}{Planck Collaboration, 2020}), the mass of the black hole is $M_{\rm BH} \approx 10^{23} \,$g. This means that for black holes that are still evaporating today or have already evaporated ($M\lesssim10^{15}\,$g), more neutrinos are emitted than photons. Black holes of higher masses have too low temperature, and observation of their Hawking radiation is inaccessible to modern instruments (see Tab.~\ref{tab:BH_examples}). This is what makes it possible to use neutrinos to set tighter constraints on the prevalence of primordial black holes than other constraints on the observation of electromagnetic radiation.

Fig.~\ref{fig:main} shows the total neutrino emission spectra from primordial black holes of different monochromatic mass distributions in the range from $10^{9.2}$\,g to $10^{11.3}$\,g in logarithmic steps of 0.1. Black holes of smaller masses evaporate in time $\lesssim1$\,sec, and all neutrino emission in this time interval will be in thermodynamic equilibrium and cannot be observed. At masses $> 10^{11.1}$\,g, the neutrino emission of primordial black holes does not exceed the background emission of other sources. For comparison, the Grand Unified Neutrino Spectrum -- GUNS is also given (for details see \hyperlink{refs:7}{Vitagliano et al., 2020}; \hyperlink{refs:10}{Ivanchik et al., 2024b}) -- the total observed or theoretically calculated neutrino flux emitted due to various astrophysical phenomena, both cosmological and local. From the plot presented, it can be seen that the neutrino emission of primordial black holes exceeds the neutrino background in the energy range $[10^{-2}\div 1]$\,eV. It is worth noting that accounting for neutrino mass can affect the behavior of the spectra, since massive neutrinos can become non-relativistic and be trapped by the gravitational potential of the Galaxy. However, this effect should not affect the energy range $[10^{-2}\div 1]$\,eV.

\begin{figure}[H]
\begin{center}
        \includegraphics[width = 0.5\textwidth]{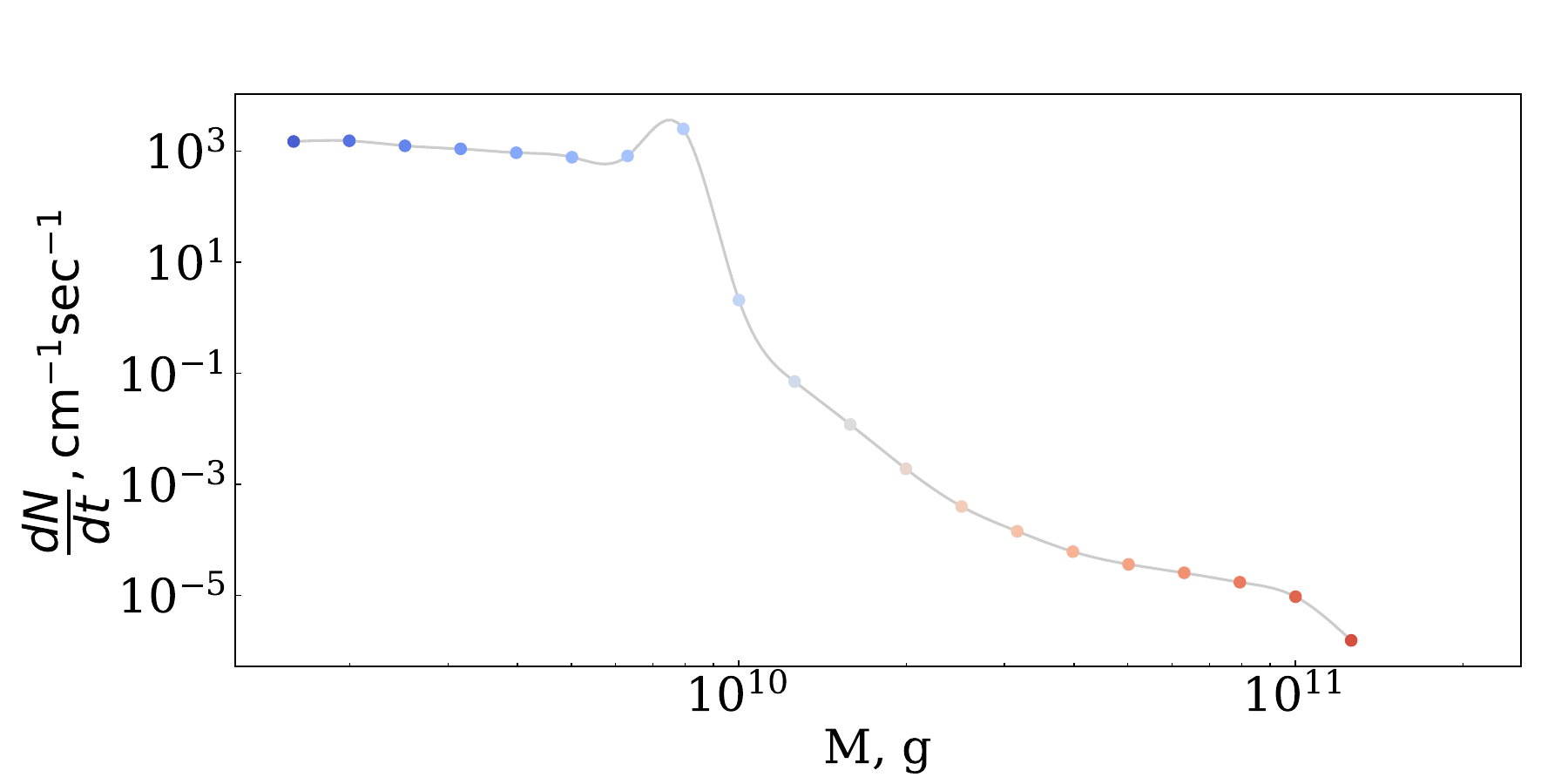}
        \caption{The total flux of neutrinos of all energies exceeding the background. The colors of the dots correspond to the colors of the curves in Fig. \ref{fig:main}. The gray curve shows spline interpolation.}
        \label{fig:area}
\end{center}
\end{figure}

By integrating by energy the difference between the neutrino flux density from primordial black holes and the background emission in the energy range where this difference is positive (for example, the dashed gray lines indicate the range of integration for the spectrum of black holes with mass $10^{9.2}$\,g), one can obtain the neutrino flux exceeding the background. This is presented in Fig.~\ref{fig:area}, from where one can see that the maximum excess over the background corresponds to the mass range $[10^{9.2};10^{9.9}]\,$g.

\section*{Conclusion}
\noindent

The following conclusions were obtained as a result of the study:

\begin{enumerate}
    \item The Hawking radiation of black holes with masses $\lesssim 10^{23}\ $ g is dominated by neutrino radiation. Due to this, the Hawking radiation of primordial black holes can potentially be observed in the neutrino component, while being undetectable in the electromagnetic radiation.
    \item All black holes of practical interest for observing Hawking radiation ($M_{\rm BH} \lesssim 10^{15}\, $g) will evaporate predominantly into the neutrino component.
    \item There is a potential opportunity to detect the neutrino Hawking radiation signal of primordial black holes in the energy range $[10^{-2}\div 10^1]$\,eV. At higher energies the radiation will be under the total background of other astrophysical neutrino sources.  
\end{enumerate}

Currently, the energy range $[10^{-2}\div 10^1]$\,eV is not an observational target of existing or planned neutrino experiments. The \hyperlink{refs:17}{PTOLEMY Collaboration, 2019} plans to study cosmological neutrinos at lower energies ($\sim 10^{-3}$ eV), while the experiments of the \hyperlink{refs:13}{JUNO Collaboration, 2016}, the \hyperlink{refs:20}{Hyper-Kamiokande Proto-Collaboration,  2018}, and the \hyperlink{refs:14}{DUNE Collaboration, 2020}, target significantly higher energy neutrinos ($\gtrsim 1$\ MeV). However, despite the lack of suitable instruments, the presented theoretical prediction may stimulate the development of neutrino detection techniques in this energy range as well. If the expected signal is not present in the observations, the current constraints on the prevalence of primordial black holes could be significantly improved. If the observations agree with the predicted neutrino emission, this could be the first case of detecting Hawking radiation, which would confirm its existence. In addition, observations will allow to judge the initial mass of primordial black holes, the time of their formation, which in turn will open new opportunities to look into even earlier moments of the existence of the universe than those that are available at the moment.

As a further development of research on this topic, it is planned to study the neutrino spectra of primordial black holes of extended mass spectra, to evaluate the effect of non-zero neutrino mass on the emission spectrum, and to include in the analysis the interaction of neutrinos with their surroundings as they propagate through the Universe.

\section*{Acknowledgments}

The authors are grateful to the reviewers for careful reading and useful comments that allowed to improve the text of this paper. The work was supported by the RSF grant № 23-12-00166. 

\section*{References}

\begin{enumerate}
    \item \hypertarget{refs:1}{A.-C. Eilers, R.A. Simcoe, M. Yue et al.}, \href{https://arxiv.org/pdf/2211.16261}{The Astrophysical Journal, Volume 950, Issue 1, id.68, 9 pp. (2023).} 
    \item \hypertarget{refs:2}{A. Arbey, J. Auffinger}, \href{https://arxiv.org/pdf/1905.04268}{The European Physical Journal C, Volume 79, Issue 8, article id. 693, 26 pp (2019).} 
    \item \hypertarget{refs:3}{A. Arbey, J. Auffinger}, \href{https://arxiv.org/pdf/2108.02737}{The European Physical Journal C, Volume 81, Issue 10, article id.910 (2021).} 
    \item \hypertarget{refs:4}{V.V. Barinov, B.T. Cleveland, S.N. Danshin et al.}, \href{https://arxiv.org/pdf/2109.11482}{Physical Review Letters, Volume 128, Issue 23, article id.232501 (2022).} 
    \item \hypertarget{refs:5}{N. Bernal, V. Mu$\tilde{{\rm n}}$oz-Albornoz, S. Palomares-Ruiz et al.}, \href{https://arxiv.org/pdf/2203.14979}{Journal of Cosmology and Astroparticle Physics, Volume 2022, Issue 10, id.068, 37 pp. (2022).} 
    \item \hypertarget{refs:7}{E. Vitagliano, I. Tamborra, G. Raffelt}, \href{https://arxiv.org/pdf/1910.11878}{Reviews of Modern Physics, Volume 92, Issue 4, article id.045006 (2020).} 
    \item \hypertarget{refs:8}{A.D. Dolgov, N.A. Pozdnyakov}, \href{https://arxiv.org/pdf/2107.08231}{Physical Review D, Volume 104, Issue 8, article id.083524 (2021).} 
    \item \hypertarget{refs:6}{A. Dolgov, J. Silk}, \href{https://journals.aps.org/prd/abstract/10.1103/PhysRevD.47.4244}{Physical Review D, Volume 47, Issue 10, pp.4244-4255 (1993).} 
    \item \hypertarget{refs:9}{A.V. Ivanchik, O.A. Kurichin, V.Yu. Yurchenko}, \href{https://link.springer.com/article/10.1007/s11141-024-10324-9}{Radiophysics and Quantum Electronics, Volume 66, Issue 9, pp. 639-649 (2024a).} 
    \item \hypertarget{refs:10}{A.V. Ivanchik, O.A. Kurichin, V.Yu. Yurchenko}, \href{https://arxiv.org/pdf/2404.07081}{Universe, Volume 10, Issue 4, id.169 (2024b).} 
    \item \hypertarget{refs:11}{B.J. Carr, K. Kohri, Y. Sendouda et al.}, \href{https://arxiv.org/pdf/0912.5297}{Physical Review D, vol. 81, Issue 10, id. 104019 (2010).} 
    \item \hypertarget{refs:12}{B.J. Carr, K. Kohri, Y. Sendouda et al.}, \href{https://arxiv.org/pdf/2002.12778}{Reports on Progress in Physics, Volume 84, Issue 11, id.116902, 53 pp. (2021).} 
    \item \hypertarget{refs:14}{DUNE Collaboration: B. Abi, R. Acciarri, M.A. Acero et al.}, \href{https://arxiv.org/pdf/2002.02967}{Journal of Instrumentation, Volume 15, Issue 08, pp. T08008 (2020).} 
    \item \hypertarget{refs:13}{JUNO Collaboration: F. An, G. An, Q. An et al.}, \href{https://arxiv.org/pdf/1507.05613}{Journal of Physics G: Nuclear and Particle Physics, Volume 43, Issue 3, article id. 030401 (2016).} 
    \item \hypertarget{refs:15}{Planck Collaboration: N. Aghanim, Y. Akrami, M. Ashdown et al.}, \href{https://arxiv.org/pdf/1807.06209}{Astronomy \& Astrophysics, Volume 641, id.A6, 67 pp. (2020).} 
    \item \hypertarget{refs:16}{PROSPECT Collaboration: M. Andriamirado, A.B. Balantekin, C.D. Bass et al.}, \href{https://arxiv.org/pdf/2406.10408}{eprint arXiv:2406.10408 (2024).} 
    \item \hypertarget{refs:17}{PTOLEMY Collaboration: M.G. Betti, M. Biasotti, A. Bosc\'{a} et al.}, \href{https://arxiv.org/pdf/1902.05508}{Journal of Cosmology and Astroparticle Physics, Issue 07, article id. 047 (2019).} 
    \item \hypertarget{refs:18}{STEREO Collaboration: H. Almaz\'an, L. Bernard, A. Blanchet et al.}, \href{https://arxiv.org/pdf/2210.07664}{Nature, Volume 613, Issue 7943, p.257-261 (2023).} 
    \item \hypertarget{refs:19}{LIGO Scientific Collaboration: R. Abbott, T.D. Abbott, F. Acernese et al.}, \href{https://arxiv.org/pdf/0711.3041}{Physical Review D, Volume 109, Issue 2, article id.022001 (2024).}  
    \item \hypertarget{refs:21a}{I.D. Novikov, A.G. Polnarev, A.A. Starobinsky et al.}, \href{https://articles.adsabs.harvard.edu/pdf/1979A%26A....80..104N}{Astronomy and Astrophysics, vol. 80, no. 1, p. 104-109 (1979).} 
    \item \hypertarget{refs:20}{Hyper-Kamiokande Proto-Collaboration: K. Abe, Ke. Abe, H. Aihara et al.}, \href{https://arxiv.org/pdf/1805.04163}{eprint arXiv:1805.04163 (2018).} 
    \item \hypertarget{refs:21}{D.N. Page}, \href{https://journals.aps.org/prd/abstract/10.1103/PhysRevD.13.198}{Physical Review D, Volume 13, Issue 2, pp.198-206 (1976a).} 
    \item \hypertarget{refs:22}{D.N. Page}, \href{https://journals.aps.org/prd/abstract/10.1103/PhysRevD.14.3260}{Physical Review D (Particles and Fields), Volume 14, Issue 12, pp.3260-3273 (1976b).} 
    \item \hypertarget{refs:23}{A.P. Serebrov, R.M. Samoilov, V.G. Ivochkin et al.}, \href{https://journals.aps.org/prd/abstract/10.1103/PhysRevD.104.032003}{Physical Review D, Volume 104, Issue 3, article id.032003 (2021).} 
    \item \hypertarget{refs:24}{D. Harlow}, \href{https://arxiv.org/pdf/1409.1231}{Reviews of Modern Physics, Volume 88, Issue 1, id.015002 (2016).} 
    \item \hypertarget{refs:25}{T. Sj\"{o}strand, S. Ask, J.R. Christiansen et al.}, \href{https://arxiv.org/pdf/1410.3012}{Computer Physics Communications, Volume 191, p. 159-177 (2015).} 
    \item \hypertarget{refs:26}{S. Hawking}, \href{https://articles.adsabs.harvard.edu/pdf/1971MNRAS.152...75H}{Monthly Notices of the Royal Astronomical Society, Vol. 152, p. 75, (1971).} 
    \item \hypertarget{refs:27}{S.W. Hawking}, \href{https://www.nature.com/articles/248030a0}{\nat, Volume 248, Issue 5443, pp. 30-31 (1974).}  
    \item \hypertarget{refs:28}{S.W. Hawking}, \href{https://link.springer.com/article/10.1007/BF02345020}{Communications In Mathematical Physics, Volume 43, Issue 3, pp.199-220 (1975).} 
    \item \hypertarget{refs:29}{G.F. Chapline}, \href{https://www.nature.com/articles/253251a0}{Nature, Volume 253, Issue 5489, pp. 251-252 (1975).} 
    \item \hypertarget{refs:30}{P.A. Chernikov, A.V. Ivanchik}, \href{https://arxiv.org/pdf/2302.05251}{Astronomy Letters, Volume 48, Issue 12, p.689-701 (2022).} 
    \item \hypertarget{refs:31}{J. Yang, F. Wang, X. Fan et al.}, \href{https://arxiv.org/pdf/2109.13942}{The Astrophysical Journal, Volume 923, Issue 2, id.262, 22 pp. (2021).} 
\end{enumerate}

\end{document}